# Using CamiTK for Rapid Prototyping of Interactive Computer Assisted Medical Intervention Applications*

Emmanuel Promayon, Céline Fouard, Mathieu Bailet, Aurélien Deram, Gaëlle Fiard, Nikolai Hungr, Vincent Luboz, Yohan Payan, Johan Sarrazin, Nicolas Saubat, Sonia Yuki Selmi, Sandrine Voros, Philippe Cinquin and Jocelyne Troccaz

*Abstract—* Computer Assisted Medical Intervention (CAMI hereafter) is a complex multi-disciplinary field. CAMI research requires the collaboration of experts in several fields as diverse as medicine, computer science, mathematics, instrumentation, signal processing, mechanics, modeling, automatics, optics, etc.

CamiTK[1] is a modular framework that helps researchers and clinicians to collaborate together in order to prototype CAMI applications by regrouping the knowledge and expertise from each discipline. It is an open-source, cross-platform generic and modular tool written in C++ which can handle medical images, surgical navigation, biomedicals simulations and robot control.

This paper presents the Computer Assisted Medical Intervention ToolKit (CamiTK) and how it is used in various applications in our research team.

## I. INTRODUCTION

The field of CAMI emerged about 30 years ago from the converging evolution of medicine, physics, materials, electronics, computer science and robotics. CAMI aims at providing tools that allow the clinician to use multi-modal data in a rational and quantitative way in order to plan, simulate and accurately and safely execute mini-invasive medical interventions. Medical interventions include both diagnostic and therapeutic acts.

CAMI can be described as a perception – decision – action loop (Fig.1).

The perception phase includes data acquisition and processing, the development of sensors and their calibration. Data can come from medical imaging devices (CT, MR, US for instance) or from specific sensors (such as localizers that track instruments or biomechanical sensors for tissue characterization); they may also be a priori knowledge embedded in models (organ shape statistics, atlases) or in the representation of a clinical protocol. This information is used in the decision phase for planning and/or simulation of a medical intervention. Finally, the action phase is implemented using guiding systems (navigational assistance or robots) to perform the intervention as planned. As can be seen in the following sections, CamiTK [1] can be used at any stage of the loop.

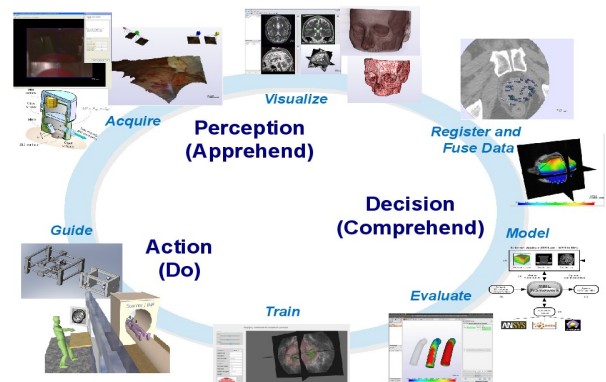

Figure 1. Using CamiTK in the CAMI perception – decision – action loop.

### A) Related Work

Numerous programming frameworks exist, generally focusing on one or two fields of CAMI. For example, some of the most popular free and open-source libraries or applications for scientific visualization [2] are Vtk[2], Paraview[2], VolView[2], as well as proprietary software such as Amira. In the field of medical imaging and medical image processing, one can refer to the following open-source applications: 3DSlicer[2], Itk-snap[3], Osirix[4], GIMIAS[5], MedINRIA[6], MITK[7], CreaTools[8], as well as the proprietary sofware Analyze[9] and Matlab. Numerous platforms also exist for biomechanical modeling and simulation, for medical robotics, or even for image-guided surgery applications such as IGSTK[2]. Although prototyping CAMI applications can be done by using two or more applications (see for example [3]), to our knowledge, none of these platforms were originally designed in order to integrate *all* the CAMI research fields in one single framework. Our initial motivation was to design a C++ open-source, cross-platform generic framework in order to provide a complete and unique development kit to test and validate an entire interactive CAMI application from perception and virtual patient building to actual intraoperative guidance.

*This work was supported by French state funds managed by the ANR within the Investissements d'Avenir programme (Labex CAMI) under reference ANR-11-LABX-0004, IBISA, the région Rhône-Alpes, Carnot Institute LSI and ECCAMI.

All authors are with UJF-Grenoble 1 / CNRS / TIMC-IMAG UMR 5525, Grenoble, F-38041, France;

e-mail: GivenName.FamilyName@imag.fr.

[1] http://camitk.imag.fr

[2] http://www.kitware.com/opensource/provensolutions.html
[3] http://www.itksnap.org
[4] http://www.osirix-viewer.com
[5] http://www.gimias.org
[6] http://medinria.com
[7] http://www.mitk.org
[8] http://www.creatis.insa-lyon.fr/site/en/CreaTools_home
[9] http://analyzedirect.com

## II. THE CamiTK PROJECT

The goal of the CamiTK project is to gather not only the knowledge of the specialists from each area of the CAMI field, but also to gather their know-how by facilitating data exchange, software prototyping of applications and thus leading faster and more efficiently toward clinical validation and benefit to the patient. CamiTK's general design is inspired by the Component-Based Software Engineering (CBSE) concept and therefore encourages reusing rather than reinventing the wheel. Rapid prototyping of CAMI applications is made easy by assembling pre-built CamiTK components (called CamiTK extensions) instead of continuously patching onto existing code.

CamiTK is composed of the CamiTK core and CamiTK extensions. The CamiTK core defines and implements all the CBSE concepts and architecture. Each extension (i.e., plugin), adds a new functionality to the framework. There are four types of extensions:

- component extensions provide data I/O features,
- action extensions provide data processing features,
- viewer extensions provide interaction and visualization features, and
- application extensions provide a way to design a GUI main window.

CamiTK is available through its website http://camitk.imag.fr where you will find information about the application, technical documentation, a software forge and a collaborative wiki about the modular framework.

### A) Application and Plugins

Although one can compile CamiTK from source, a user package is also available for different platforms (Windows and Linux Debian/Ubuntu packages). It features the flagship application CamiTK-imp with several basic plugins. CamiTK-imp is an all-in-one application that allows users to load data and interact with them. It grants users with the main CAMI functionalities: data visualization (load 2D/3D medical images), image analysis (interactions on the image) and biomechanical simulations.

Provided component extensions include I/O support for several formats of volume image and mesh management (e.g., DICOM, Itk, Vtk, Analyze images and Vtk, VRML, Obj, msh, PML geometries). Provided viewer extensions include a classical medical image widget (including axial, coronal, sagittal plane display, 3D and arbitrary slice views). Provided action extensions include volume rendering, volume image filtering algorithms using ITK, mesh processing and 3D reconstruction, and biomechanical model simulation.

CamiTK-imp also allows users to load any supplementary CamiTK extensions, developed by other experts in order to extend the functionalities of the framework.

### B) Software Architecture

Extending CamiTK is easily possible by adding a new extension. An expert from a specific CAMI field can add his or her own contribution by developing his or her own extension. The open-source LGPL v3 license used for the framework leaves the extension's licensing options up to the expert.

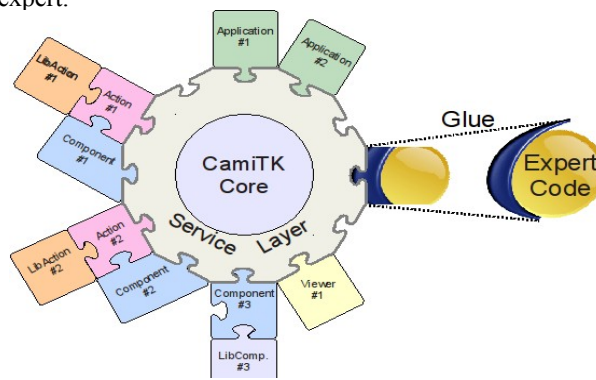

Figure 2. CamiTK Framework Overview: everything is an extension.

In order to be modular and relevant to any CAMI field, CamiTK core architecture distinguishes between domain logic (data I/O and data processing) and application logic (data viewing/interaction and user interaction). CamiTK offers an implementation of most current application logic and default domain logic behaviours. As domain logic is specific to a research field, the specialist can focus only on his or her area of expertise by developing his or her specific CamiTK component or action extension. Application logic is mostly based on Qt for the GUI and VTK for scientific visualization, two very popular open-source libraries.

CamiTK also provides a service layer which generalizes concepts and predefines generic behaviors. It allows for automatic handling of the communication between application logic and domain logic. Therefore, the specialist only needs to encapsulate/wrap the domain-specific code into a CamiTK extension in order to plug it into the service layer (see Fig. 2).

To simplify the extension process, a CamiTK-wizard application assists the expert in creating his or her own extension by automatically generating the source code skeleton and platform specific build environment.

## III. EXAMPLES OF APPLICATIONS

This section provides examples of projects developed under CamiTK and that reach the three areas of CAMI: Perception (data acquisition from patients), Decision (image processing, data fusion, biomechanical modeling) and Action (medical simulators and robotic guidance) to illustrate the multidisciplinary possibilities of CamiTK. Note that some of the mentioned extensions are not available under an open-source license.

### A. Perception: Video Capture

The distributed endoscopy project [5] aims at providing the surgeon with a large view of his field of operation during endoscopic surgery. To build a 3D view of the patient's abdomen, small sets of endoscopic cameras are inserted into small incisions. Each stereoscopic camera pair is handled by a specific CamiTK component extension. Several CamiTK action extensions are used to calibrate the cameras, reconstruct the surface of the inner organ in 3D and to visualize the resulting textured 3D mesh (Fig. 3).

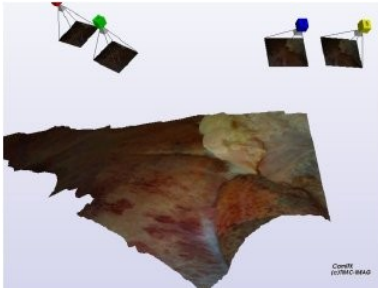

Figure 3. Endoscopic reconstruction of porcine organs with two pairs of stereo video cameras.

### B. Perception: Tissue Characterization

Estimating soft tissue elasticity helps improve the accuracy of biomechanical simulations and consequently enables the designing of better models to help planning and intra-operative assisted surgery. LASTIC (for Light Aspiration device for in vivo Soft TIssue Characterization) is based on the pipette aspiration principle [6] and aims at *in vivo* characterization of the elastic modulus of soft tissues. Once placed on the studied soft tissues, this 33x34mm cylinder creates a negative pressure which deforms the tissues surface. The height of this deformation is measured on the images captured during the acquisition for several negative pressures (see Fig. 4). The corresponding height/negative pressure curve allows the elastic modulus of the tissues to be estimated using an inverse method. This elastic modulus can then be integrated in a patient-specific model and accurately simulate the behavior of the tissues.

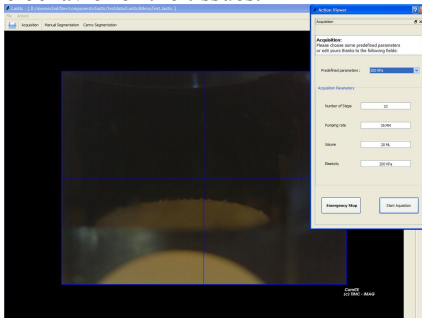

Figure 4. Specific CamiTK application for tissue characterization.

A specific main window CamiTK extension (see Fig. 4) has been built to handle the communication and acquisition synchronization via USB ports between the programmable syringe pump that create the negative pressure, the manometer that measures this pressure, and the LASTIC mini camera that acquires the images of the deformed soft tissues at each step. Furthermore, the segmentation of the deformation height is computed using CamiTK image processing action extensions. Finally, the elastic modulus estimation is evaluated through an inverse method, based on a library of precomputed deformations, programmed as another action extension. The GUI provided by CamiTK allows an intuitive manipulation of LASTIC and it has already been used on the tongue, brain, skin, and foot.

### D. Decision: Information Fusion

Prostate brachytherapy, using low dose rate radioactive seeds, is a common treatment for the management of localized prostate cancer. The Dorgipro project aims at determining the impact of the true location and orientation of the seed on the treatment, as opposed to the theoretical, planned location actually used in conventional brachytherapy. Component extensions are used to handle patient CT scans and action extensions to precisely determine the location and orientation of the actual inserted seeds. Then, seeds are reconstructed and the dosimetry is performed and superimposed to the original image to observe the actual irradiation of the tumor and surrounding tissues (see Fig. 5).

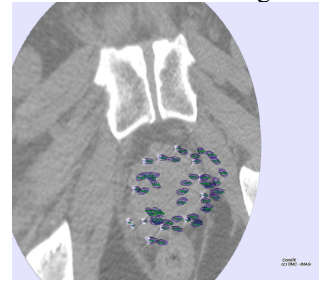

Figure 5. Comparison of the theoretical and actual position and orientation of radioactive seeds in the treatment of localized prostate cancer.

### C. Decision: Biomechanical Modeling

CamiTK can run, evaluate and compare biomechanical models through the MML framework [6]. Models can be described in a generic way in XML files (MML for monitoring, PML for physical models and LML for loads). A CamiTK component allows simulations to be run using a variety of software (ANSYS[10], ArtiSynth[11], Sofa[12]) and compare the results with real experiments or other simulations in a transparent way for the user. A specific CamiTK application is used for calculating statistics (e.g., RMS error, standard deviation …) on metrics generated by the simulations (computing time, displacements, geometric deviation...) allowing for fast model and parameter space evaluation even when hundreds of simulations are to be evaluated.

An in-house biomechanical modeling and C++ simulation library is wrapped as a MML plugin available through a specific action extension. It provides static and dynamic analysis of mechanical models, solid and shell finite element models and supports various linear and non-linear hyperelastic materials (Hookean, Neo Hookean, Saint-Venant), see Fig. 6.

---

[10] http://www.ansys.com
[11] http://www.magic.ubc.ca/artisynth/pmwiki.php
[12] http://www.sofa-framework.org

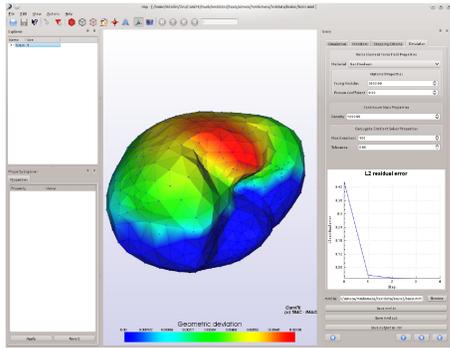

Figure 6. Brain-shift static simulation using tetrahedrons and a Neo-Hookean material (in CamiTK-imp).

*E. Decision and Action: Biopsy Simulator*

Prostate biopsy procedures consist of the removal of small prostate tissues for cancer diagnosis. The limits of the current training methods, based primarily on companionship, have lead to the development of a simulator which provides a virtual environment for transrectal ultrasound (TRUS) guided biopsies. The simulator is composed of a specific main window extension and a Phantom/Omni haptic device (Sensable, Wilmington, MA, USA) that tracks the virtual probe motion. The purpose is to provide feedback to the user by displaying the precise position of biopsy samples on a 3D viewer. The simulator provides various exercises, allowing the user to train on multiple aspects of a TRUS exam. In this application, a main window application extension is used for data visualization and data management. Open-source component extensions have been developed to load ultrasound and MRI images (Analyze format) and prostate meshes (OFF format). Biopsy results are saved by another component extension that manages the patient images, mesh and biopsy localization all in the same reference frame. CamiTK facilitates the analysis of the results in order to assess the impacts of the simulator on physician training.

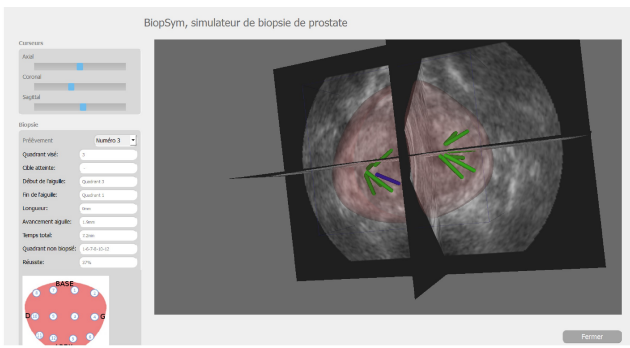

Figure 7. Biopsy Simulator. 3D view showing a biopsy result (ultrasound image, registred prostate mesh and biopsy localization).

*F. Action: Robot Control*

ROBACUS is an interventional radiology robot for CT and MRI-guided needle interventions. It makes use of perception and reasoning to apply a diagnostic or therapeutic act on a patient in the form of a needle puncture. The goal of the project is to allow a radiologist to position and insert a needle directly inside the imager's tunnel with higher accuracy and less iteration compared to the conventional manual techniques. The robot is calibrated to the image through a pair of fiducials, allowing potentially complex trajectories that would be extremely challenging to reproduce manually. The CamiTK environment allows the project to combine the existing medical image viewer and various image analysis algorithms for the segmentation and registration of the fiducials with actions and components specifically written to control the robot. In addition, a state machine designer (see Fig. 8) is used to group the various CamiTK actions in order to create the GUI for the experimental protocol.

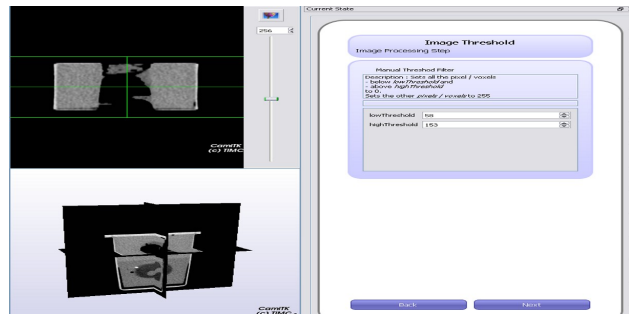

Figure 8. The CamiTK state-machine designer can be used to model a clinical protocol.

IV. CONCLUSION

Rapid prototyping of CAMI integrated applications is made easy by CamiTK thanks to its modular architecture. This paper presented several projects based on CamiTK, all centered around the CAMI preception-decision-action loop. The aim of CamiTK is to grow with the research community needs and participations. Thus, we encourage the reader to check the website and submit comments, suggestions and questions. The roots of the modular CamiTK design go back 10 years, which ensures its flexibility and adaptability.